# Superconducting Properties of Combustion Synthesized $MgB_2$

Yoshihiko Takano, Nobutaka Oguro, Yoshinari Kaieda, and Kazumasa Togano

*National Institute for Materials Science, 1-2-1 Sengen, Tsukuba 305-0047 Japan*

We have successfully prepared the $MgB_2$ superconducting bulk and powdered materials by the method of combustion synthesis. The starting materials used in this study were powders of Mg and B. X-ray powder diffraction pattern was well assigned to the P6/mmm $MgB_2$ phase. The temperature dependence of magnetization shows sharp superconducting transition around 38K. The critical current density can be estimated from the hysteresis of the magnetization curvature using the Bean's model. The powdered sample shows a high critical current density of $2x10^6$ A/cm$^2$ at 5K under the magnetic field of 1T.

Corresponding author,
Name: Yoshihiko Takano
Email: TAKANO.Yoshihiko@nims.go.jp

------------------------------------------------------------------------

## Introduction

The $MgB_2$ superconductor was discovered by Nagamasu et al. in the very beginning of the 21st century in January 2001 [1]. Since it has the highest superconducting transition temperature (Tc~39) in the metallic superconductors, its superconducting properties have been investigated to elucidate the mechanism of superconductivity [2-10]. $MgB_2$ is expected to be a candidate material for superconducting applications, because it shows the relatively high upper critical fields and no weak coupling of grains [11-14]. The $MgB_2$ superconducting wires can be fabricated by the powder-in-tube method, and show fairly high critical current densities. To improve their properties, high quality $MgB_2$ powder is needed. However, it is difficult to obtain high purity $MgB_2$ powder because Mg is extremely volatile, and the difference of melting temperature between Mg and B is more than 1500C. Especially for the application use, simple and high-speed production technique to obtain high purity $MgB_2$ powder is required.

Recently, we have successfully prepared $MgB_2$ superconductor using the combustion synthesis. The combustion synthesis method has a great advantage for application in that its reaction time is short and its process is simple. In this paper, we report the superconducting properties of $MgB_2$ bulk and powder samples prepared by the combustion synthesis.

## Experimental

The schematic diagram of combustion synthesis is shown in fig. 1. The starting materials used in this study were powder Mg (99.9%, <500um, Yamaishi-Kinzoku) and B (96.06%, 0.89um, Hermann C. Starck). The appropriate mixture of starting materials, Mg and B, were pressed by cold-isostatic-pressing (CIP) under 300 MPa. And the obtained Mg+2B pellets were surrounded by Ti+B or Ti+C and placed in the crucible. Since the heat of reaction of Mg and B was not enough for chain reaction, then Ti+B or Ti+C was used to assist the reaction. The crucible was placed in the high vacuum chamber. The combustion synthesis was started by the tungsten ignition heater in the crucible. Once its reaction started, the heat, high enough to lead to chain reaction, spread quickly through the $MgB_2$. Three $MgB_2$ samples were prepared at different conditions. The first sample named MgB2-TiB was synthesized at room temperature with Ti and B. And the sample $MgB_2$-TiC-RT and $MgB_2$-TiC-200 were made with Ti and C at room temperature and 200°C, respectively. The details of sample preparation were summarized in the previous papers [15,16].

X-ray powder diffraction was performed using Cu-Kα radiation between 20 and 80 deg. Temperature dependence of magnetization was observed from 5 to 50 K under the magnetic field of 20 Oe by superconducting quantum interference device (SQUID) magnetometer. M-H curvatures were measured under the magnetic fields between –5 and 5T. The grain size was estimated by scanning ion-beam SIM microscope using Ga ion source.

**Results and Discussion**

Figure 2 shows the optical picture of the obtained $MgB_2$-TiB sample. The sample is black, porous and soft. Its density is around 0.98g/cm$^3$, which is about one-third of the sample prepared by high pressure sintering [5]. Some of the sample was grounded by agate mortar to obtain powdered sample. The X-ray diffraction pattern of the powdered sample $MgB_2$-TB is shown in fig. 3. The obtained peaks were well assigned to the P6/mmm phase, according to the space group of the $MgB_2$ superconductor. And no other phases were detected.

The Temperature dependence of magnetic susceptibility of the bulk samples $MgB_2$-TiC-200, $MgB_2$-TiC-RT and $MgB_2$-TiB are plotted in fig. 4. The applied magnetic field was 20 Oe, measurement was performed after zero-field-cooling (ZFC) and field-cooling (FC) conditions. These three samples show sharp superconducting transition at around 38 K. The large differences between the ZFC and FC curvatures indicate that the samples have large flux pinning force. The superconducting volume fraction of $MgB_2$-TiC-RT and $MgB_2$-TiC-200 estimated from ZFC data at 5K are about 65 and 87%, respectively. On the other hand, the $MgB_2$-TiB sample shows relatively large diamagnetic signal, and its superconducting volume fraction is estimated to be about 100%. Using Ti and B for the combustion synthesis significantly improves the superconducting volume fraction.

The heat of reaction of Ti and C is about 184kJ/mol, which is considered to be insufficient to obtain high quality $MgB_2$ samples. The heat of reaction of Ti and B is about 280kJ/mol, which is 1.5 times larger than that of Ti and C. The $MgB_2$ sample of good quality was obtained by the appropriate heat given by the combustion with Ti and B. The results we obtained from $MgB_2$-TiC-200 with the starting temperature of 200°C showed the smallest superconducting volume fraction. What we think of those possible causes are as follows: 200°C was too high for the starting temperature, having made the temperature of the sample go beyond the optimum reaction range; The starting temperature, 200°C, heated the chamber itself and slowed down the cooling rate of the sample after its reaction and therefore deteriorated the quality of it. Only a short period of time is allowed for high temperature reaction of Mg due to its volatility.

Fig. 5(a) and (b) show magnetization versus magnetic field curvatures of the bulk and powdered $MgB_2$-TiB samples, respectively. These samples show characteristic curvatures of the type-II superconductor. The critical current density $J_c$ can be estimated from the hysteresis of the magnetization curvature, $\Delta M$, by Bean's model with the simple formula $J_c=30\Delta M/d$, where $d$ is the diameter of the sample [17]. The fig. 6 shows the magnetic field dependence of the critical current densities of bulk $MgB_2$-TiB sample. The $J_c$ of the bulk sample is $2.8\times10^4$ A/cm$^2$ at 5K under 1T, which are one order smaller than that of the high pressure bulk sample [5]. Since the bulk sample is porous and soft, grain connection is weak, its $J_c$ is suppressed.

The $J_c$ of the powdered $MgB_2$-TiB as a function of applied magnetic fields are plotted in fig. 7. The average particle size of the powder is around 5μm which is determined by SIM. The critical current density of the powdered $MgB_2$-TiB sample is $2\times10^6$ A/cm$^2$ at 5K under the magnetic field of 1T. This value is higher than that of the commercially available powder, Furuuchi-kagaku, [5]. It is suggested that the particles have many pinning centers in their grains.

## Conclusion

The high quality $MgB_2$ powder sample is successfully prepared by combustion synthesis using Ti and B. This method has a great advantage for application because of its short reaction time and simple process. The bulk sample showed relatively low critical current densities, since the sample was porous and has weak grain connection. On the other hand, the powdered sample shows two order high critical current densities compared to the bulk sample. The high quality powder $MgB_2$ sample prepared by combustion synthesis is useful for applications such as superconducting wires.

## References

1. J. Nagamatsu, N. Nakagawa, T. Muranaka, Y. Zenitani and J. Akimitsu, Nature **410** 63 (2001).
2. S. L. Bud'ko, G. Lapertot, C. Petrovic, C. E. Cunningham, N. Anderson, and P. C. Canfield, Phys. Rev. Lett. 86, 1877 (2001).
3. J. W. Quilty, S. Lee, A. Yamamoto, and S. Tajima, Phys. Rev. Lett. 88, 087001 (2002).
4. E. A. Yelland, J. R. Cooper, A. Carrington, N. E. Hussey, P. J. Meeson, S. Lee, A. Yamamoto, and S. Tajima, Phys. Rev. Lett. 88, 217002 (2002)
5. Y. Takano, H. Takeya, H. Fujii, H. Kumakura, T. Hatano, K. Togano, H. Kito, H. Ihara, Appl. Phys. Lett. **78**, 2914 (2001).
6. M. Xu, H. Kitazawa, Y. Takano, J. Ye, K.

Nishida, H. Abe, A. Matsushita, N. Tsujii, and G. Kido, Appl. Phys. Lett. 79, 2779(2001).
7. S. Tsuda, T. Yokoya, T. Kiss, Y. Takano, K. Togano, H. Kito, H. Ihara, and S. Shin Phys. Rev. Lett. 87, 177006 (2001)
8. S. Tsuda, T. Yokoya, Y. Takano, H. Kito, A. Matsushita, F. Yin, J. Itoh, H. Harima, and S. Shin Phys. Rev. Lett. 91, 127001 (2003)
9. A. K. Pradhan, M. Tokunaga, Z. X. Shi, Y. Takano, K. Togano, H. Kito, H. Ihara, and T. Tamegai Phys. Rev. B 65, 144513 (2002)
10. Z. X. Shi, A. K. Pradhan, M. Tokunaga, K. Yamazaki, T. Tamegai, Y. Takano, K. Togano, H. Kito, and H. Ihara Phys. Rev. B 68, 104514 (2003)
11. D.C. Larbalestier, M.O. Rikel, L.D. Cooley, A.A. Polyanskii, J.Y. Jiang, S. Patnaik, X.Y. Cai, D.M. Feldmann, A. Gurevichi, A.A. Squitieri, M.T. Naus, C.B. Eom, E.E. Hellstrom, R.J. Cava, K.A. Regan, N. Rogado, M.A. Hayward, T. He, J.S. Slusky, P. Khalihah, K. Inumaru, M. Haas, Nature 410 (2001) 186.
12. G. Grasso, A. Malagoni, C. Ferdeghini, S. Roncallo, V. Braccini, M.R. Cimberle, A.S. Siri, Appl. Phys. Lett. 79, 230 (2001).
13. H. Kumakura, Y. Takano, H. Fujii, K. Togano, Physica C 363 (2001) 179.
14. S. X. Dou, S. Soltanian, J. Horvat, X. L. Wang, S. H. Zhou, M. Ionescu, and H. K. Liu, Appl. Phys. Lett. 81, 3419 (2002)
15. Yoshinari Kaieda and Nobutaka Oguro, J. Japan Soc. of Powder and Powder Metallurgy, 49, 588 (2002)
16. Nobutaka Oguro and Yoshinari Kaieda, J. Japan Soc. of Powder and Powder Metallurgy, 49, 964 (2002)
17. C.P. Bean, Phys. Rev. Lett. 8, 250(1962).

## Figure captions

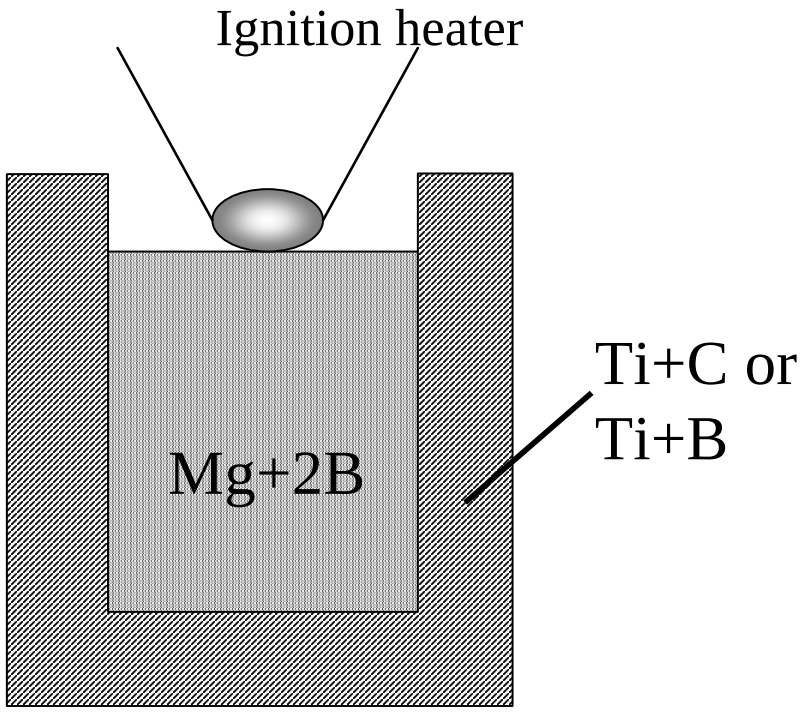

Fig 1. Schematic diagram of the combustion synthesis.

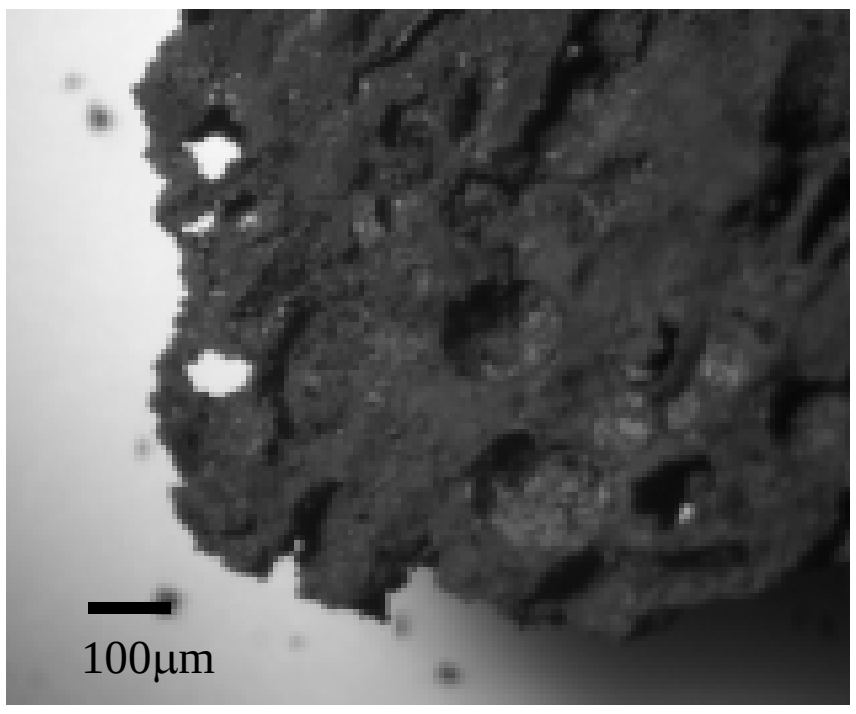

Fig 2. Optical picture of the $MgB_2$-TiB sample prepared by the combustion synthesis with Ti and B.

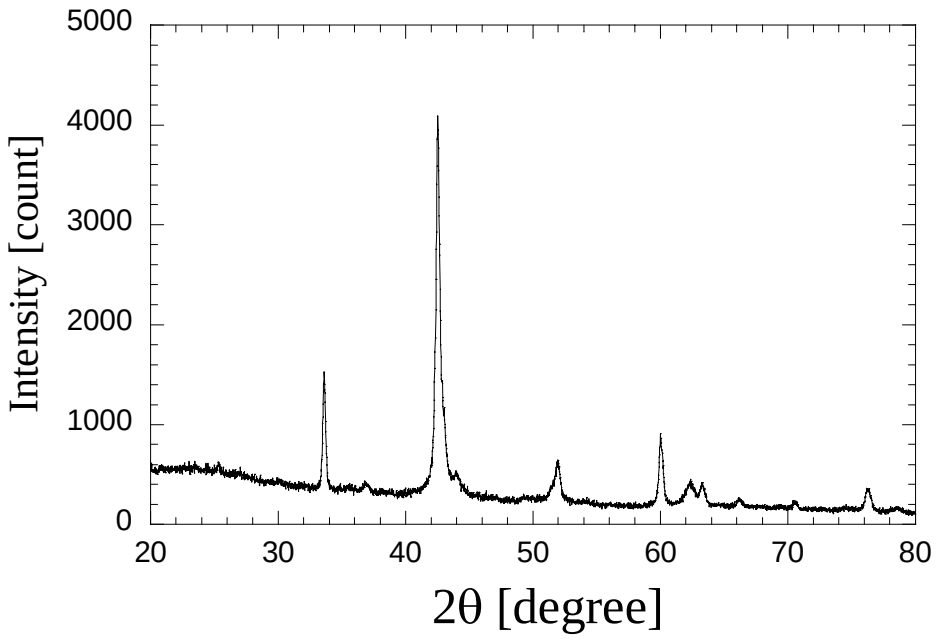

Fig 3. X-ray diffraction pattern of the powdered sample $MgB_2$-TB

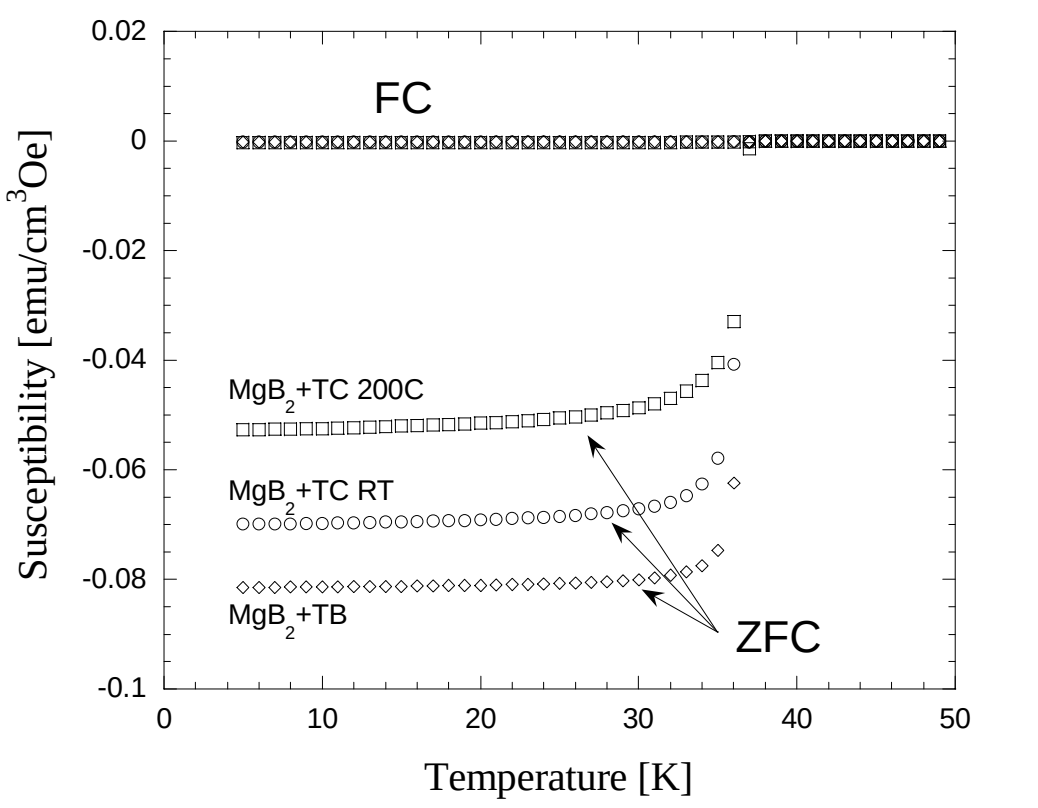

Fig 4. Temperature dependence of susceptibility of the $MgB_2$ samples. Applied magnetic field is 20 Oe.

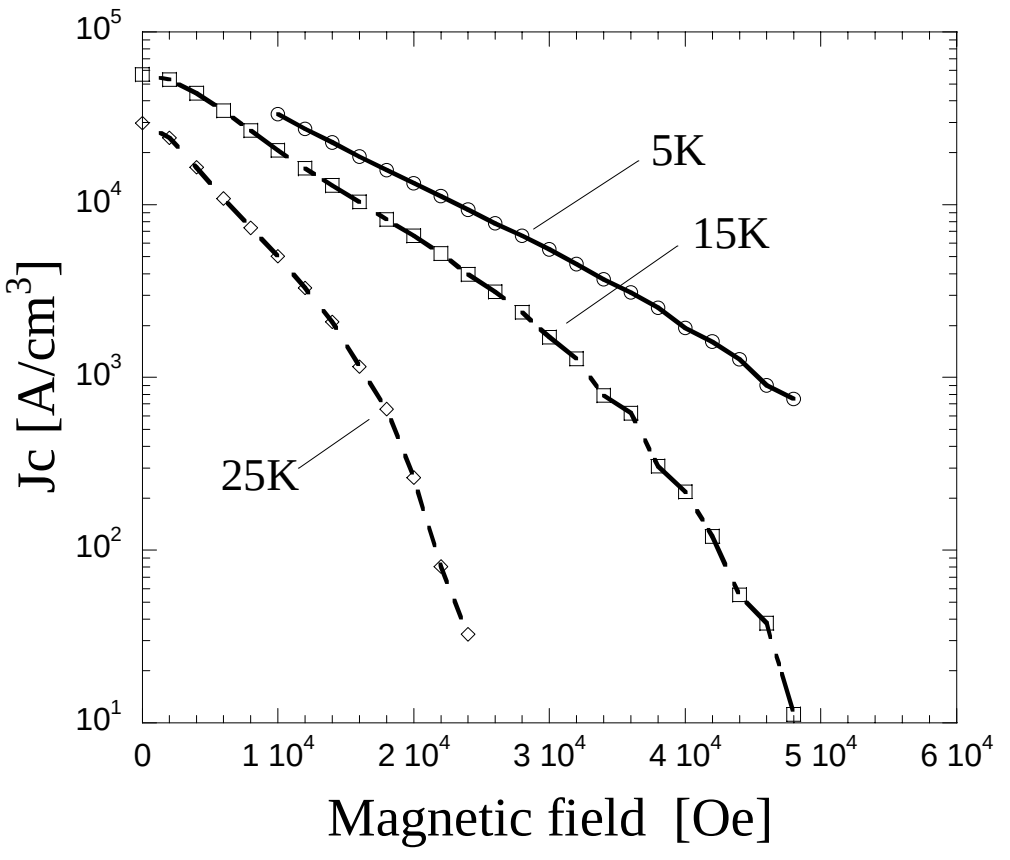

Fig 6. Critical current densities of the bulk $MgB_2$-TiB sample as a function of applied magnetic fields.

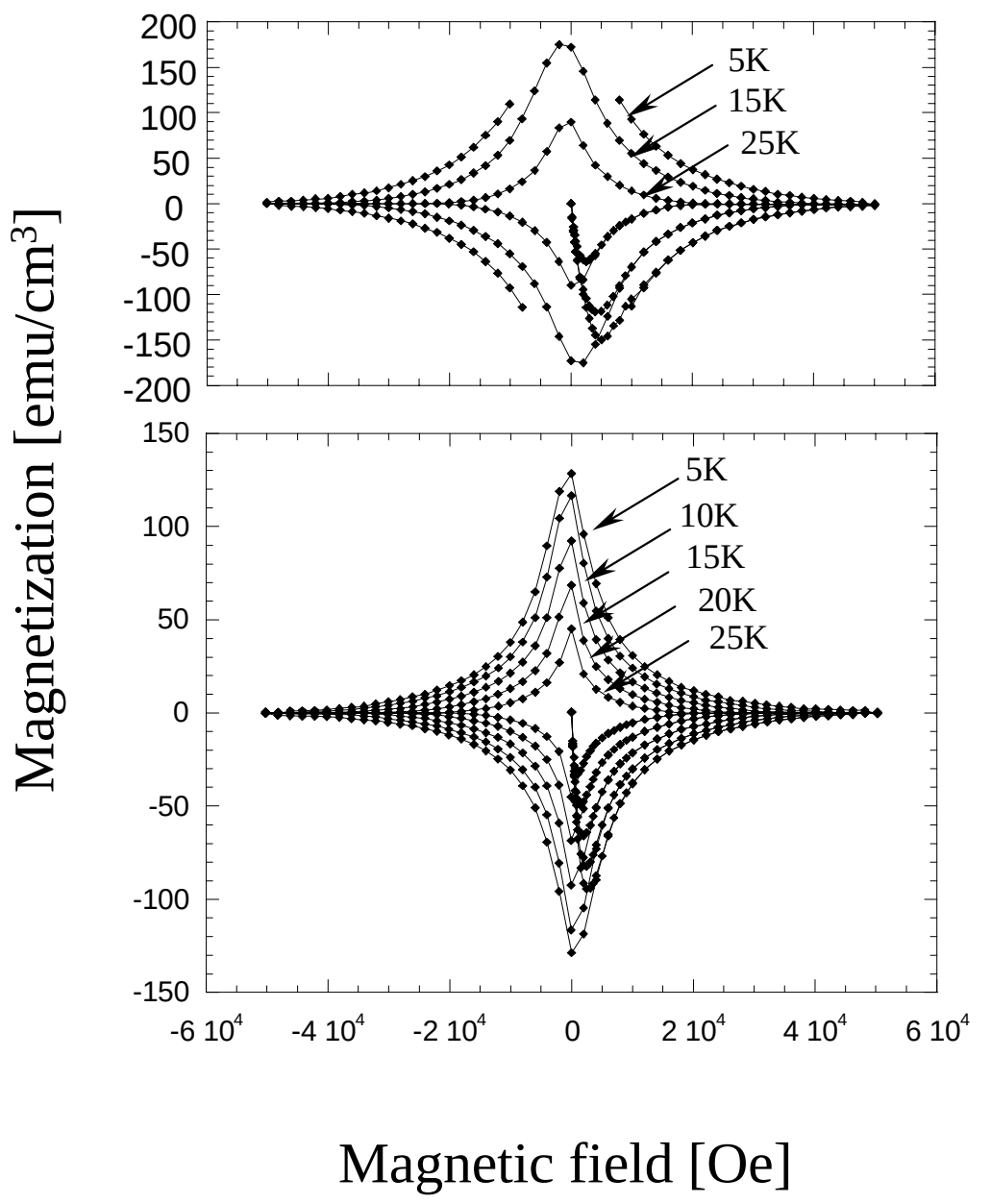

Fig 5. Magnetization versus magnetic field curvatures of the bulk and powdered $MgB_2$-TiB samples.

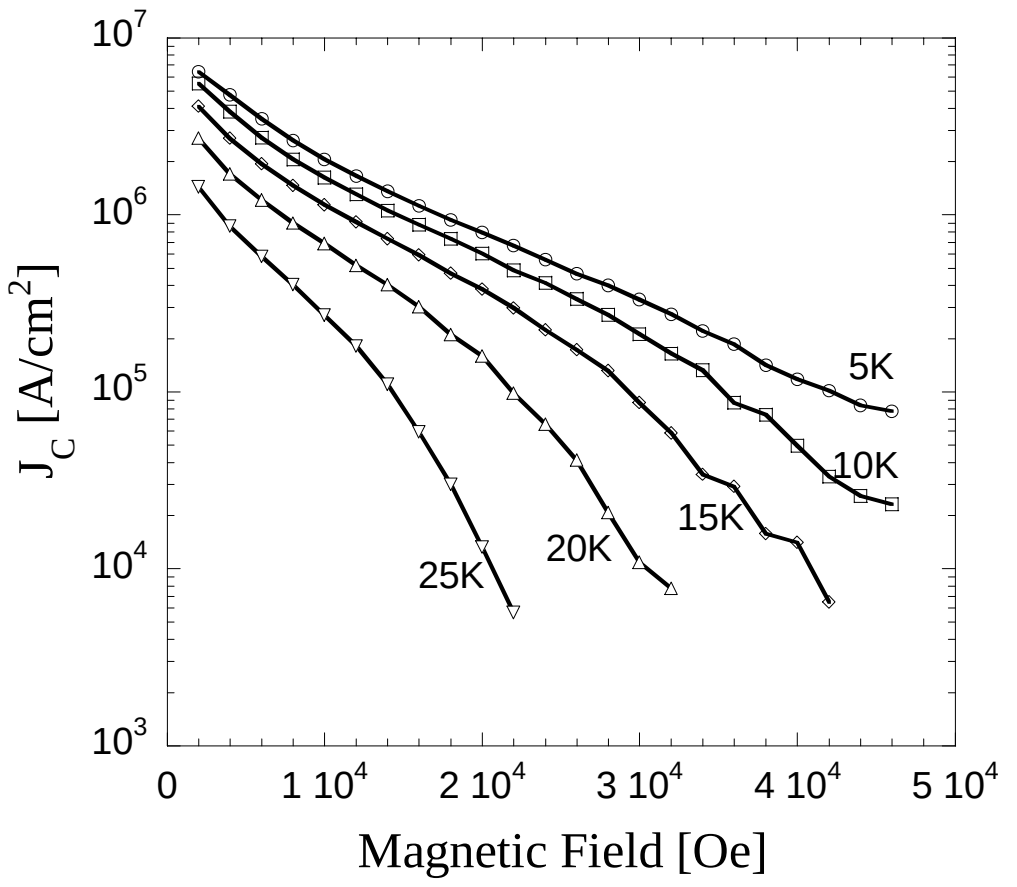

Fig 7. Critical current densities of the powdered $MgB_2$-TiB sample as a function of applied magnetic fields.